\documentclass[review,12pt]{elsarticle}

\usepackage{%amsmath,
amssymb,amsbsy,amsthm, fullpage}
\usepackage{epsfig,layout}
\usepackage{graphicx}
\usepackage{tabularx}
\usepackage{multirow}
\pdfoutput=1

\graphicspath{{images/pdf/}}

\let\epsilon\varepsilon
\let\phi\varphi

\let\epsilon\varepsilon

\newtheorem*{lemma*}{Lemma}
\newtheorem{theorem}{Theorem}

\journal{International Journal of Forecasting}
\begin{document}

%\begin{frontmatter}

\author{  Boris Ryabko$^*$ and Pavel Pristavka$^{**}$  \\ boris@ryabko.net, $ \, $ ppa@ngs.ru}
\title {
Experimental Investigation of Forecasting Methods
Based on Universal Measures 
}
%\date{}
\address{$^*$Siberian State University of Telecommunications and Information Sciences and Institute of\\  Computational Technology  of Siberian Branch of Russian Academy of Science}

\address{$^{**}$Siberian State University of Telecommunications and Information Sciences}

\begin{abstract}
We describe  and experimentally investigate a method to construct
forecasting algorithms for stationary and ergodic processes based
on universal measures (or  so-called universal data
compressors). Using  some geophysical and %%%d
economical time series as examples, we show that the precision of thus obtained
predictions is higher than that of known methods.

\end{abstract}

\maketitle

%\end{frontmatter}

\section{Introduction}

The problem of forecasting is  important for many applications.  %%%%d
 Nowadays there are many efficient forecasting methods,
which are based on different approaches and different mathematical
models (see for review, for ex., De Gooijer  \& Hyndman (2006) and Makridakis (1996)). Many  such %%%d
approaches are deeply connected with different ideas and results
of mathematical statistics,  probability theory and some other
fields.

In this paper we develop and experimentally investigate an
efficiency of prediction methods which are based on so-called
universal measures (or universal data compressors).

 By definition,  universal codes, or universal methods of lossless data compression,  are intended to 
``compress''  texts, i.e. encode them in such a way that the
length of an encoded text is shorter than the  length of the
initial one (and, of course, the original text can be recovered from %%%d
the encoded one). It is important that the text statistics is
unknown beforehand, that is why such codes are called universal.
In fact,  universal codes implicitly estimate unknown
characteristics of the processes and use them for data
compression.

A universal code is called optimal if it
encodes a
sequence of letters generated by a finite-alphabet source in such a way that the length of the encoded sequence
is asymptotically minimal.
% In turn, it is known in information theory that this minimality means that the length of the encoded sequence goes to the Shannon entropy (per letter).
First universal codes, which are  optimal for i.i.d. information sources, were discovered by Fitingoff (1966) and Kolmogorov
(1965).  The universal codes for stationary and ergodic sources with finite alphabets have been known
since 1980's; see Ryabko (1984).

 It is clear that
universal codes can be considered as a tool of  mathematical statistics and it is natural to try to apply them
for solving traditional problems of this science (like hypothesis testing, parameter estimation, etc.)
First it  was recognized in 1980's (see  Rissanen(1984) and Ryabko(1984, 1988) ) and nowadays it is shown that 
the universal codes can be efficiently used for hypothesis testing,
parameter estimation and prediction of time series with finite and real-valued alphabets; see Ryabko \& Astola (2006), Ryabko (2009).
However, there are only preliminary results which concern practical applications of prediction methods
to real data, see Ryabko \& Monarev (2005). These results were rather aimed to 
illustrate the possibility of such applications than give information about the precision of the methods.

The goal of this work is the construction, implementation and experimental estimation of the
methods of prediction suggested, which are based  on  universal codes and so-called universal measures.
To this aim, we have considered several types of time-series data: the indices of solar %%%d
activity (SA), water level, and cross-rates of some currencies. It is clear that such processes  are of great theoretical %%%d
and practical importance. For example, nowadays the statistical connections between climate and
SA are widely investigated.

The obtained experimental results show that the forecasting methods based on universal codes possess a high precision.
In principle, any universal code can be used as a basis of a prediction method, but here we use a universal
code suggested by Ryabko (1984), which have some additional useful properties, see  Krichevsky (1993), Ryabko, Astola (2006).

The outline of the paper is as follows. The next part contains necessary
definitions and some information about universal codes and measures.
The part three is  devoted to prediction of real time series and the part four is a  short conclusion.

\section{ Method and its implementation description}
\subsection{ Description of the method}
Here we will describe the forecasting method studied, %%%d
 and also provide required theoretical information.
It will be convenient at first to describe briefly the prediction problem. We consider a stationary
and ergodic source which generates sequences $x_1 x_2 ...  $ of elements (letters) from some set (alphabet)
$A$, which is either finite or real-valued. It is supposed that the probability distribution
(or distribution of limiting probabilities) $P ( x_1 = a_{i_1}, x_2 = a_{i_2}  , ... , x_t = a_{i_t} )$ (or the density
$p( x_1, x_2 , ... , x_t )$)
is unknown. Let the source generate a message $x_1, ...x_{t-1} x_t$ , $x_i \in A$   for all $i$, and the next letter
needs to be predicted.
Now let us describe the forecasting method. First we define a universal measure and then explain how it is connected with  universal codes.
By definition a measure $\mu$  is universal if for any stationary and ergodic source $P$ the following
equalities are valid:
$$
\lim_{t\rightarrow\infty} \,  t^{-1} (\log P( x_1 ...x_t ) -  ( \log \mu( x_1 ...x_t ))) = 0 ,
$$
with probability 1, and
$$
\lim_{t\rightarrow\infty} \,  t^{-1} \sum_{u \in A^t}  P(u )  \log ( P(u) / \mu( u))) = 0 .
$$ (Here and below $\log x = \log_2 x $.)
These equations show that, in a certain sense, the measure $\mu$ is an estimate of (unknown) measure $P$.
That is why the universal measures can be used for estimation of process characteristics and prediction.

 In what follows, we will directly describe a certain universal measure which will be used for prediction in this paper
 and, in principle, we
do  not have to mention  universal codes at all. %%%d
 But  the point is that  universal measures
are deeply connected with the universal codes and, if one has a universal code, one can simply convert  it into a
universal measure. It is important for practical applications, because
many universal codes are available as computer programs (so-called archivers) and, hence, they can be transformed into
universal codes and then   directly  used  for prediction,
as it is shown by Ryabko \& Monarev (2005). So, we give a short informal definition of  universal codes, whereas a full
description can be found, for ex., in  Ryabko (2010).

Roughly speaking,  a   code maps words from the set $A^t$, $t \ge 1$, into the set of words over alphabet $\{0,1\} $.
By definition, a code $U$ is universal (for the set of stationary ergodic sources),
 if for any stationary and ergodic source $P$ the following
equalities are valid:
$$
\lim_{t\rightarrow\infty} \, | U ( x_1...x_t ) | / t = H ( P)
$$
with probability 1, and
$$
\lim_{t\rightarrow\infty} \,| E_P (U ( x_1...x_t ) |)/ t = H ( P) ,
$$
where $ |v|$ is the length of the word $v$, $E_P(f)$  is the mean  of $f$ with respect  to $P$, %%%d
$H(P)$ is the Shannon entropy of $P$, i.e.
$$
H(P) = \lim_{t\rightarrow\infty} \, - \,t^{-1}   \,  \sum_{u \in A^t}  P(u )  \log ( P(u) ) ,
$$ Gallager (1968). The following statement shows that any universal code determines
a universal measure.
\begin{theorem}\label{UU}
 Let $U$ be a universal code and
$$\mu_U (w ) =  2^{-|U (w )| }/ \sum_{u \in A^{|w|}} \mu_U (u ) .
$$
Then $\mu_U$ is a universal measure.
\end{theorem}
(The simple proof of this theorem can be found  in Ryabko, (2010).)
So, we can see that, in a certain sense, the measure $\mu_U$ is a consistent (nonparametric) estimate
of the (unknown) measure $P$.

Now we a going to describe the universal measure $R$  which will be used
as a basis for forecasting in this paper. For this purpose we first describe the 
Krichevsky measure $K_m$, which is universal for the set of Markov sources of memory, or connectivity,
$m$, $m \le 0$, if $m = 0$, the source is i.i.d.  In a certain sense this measure is optimal for this set  (see for details Krichevsky (1993),
Ryabko, (2010).)
By definition,
\begin{equation}\label{km}
K_m(x_1 ... x_t) =\cases{\frac{1}{|A|^t},&if $t \leq m\,$;\cr
 & \cr
 \frac{1}{|A|^m}\: ( \frac{\Gamma(|A|/2)}{\Gamma(1/2)^{|A|}}
)^{|A|^m} \prod_{v \in A^m} \frac{\prod_{a \in A}\:
           ( \Gamma( \nu_x(v a )+ 1/2)}{( \Gamma( \bar{\nu}_x(v  )+|A|/2))}, &if $ t > m $ \, ,}
\end{equation}
where $x = x_1 ...
x_t, $ $\nu_x(v ) $
is the count of word  $v$  , occurring in the sequence $x_1 ... x_{|v|}, $
$x_2 ... x_{|v|+1}, $ $ ... , $
$x_{t - |v| + 1} ...$ $x_{t}$,
$ \bar{\nu}_x(v  )= \sum_{a \in A} \nu_x(v a ) $, $\Gamma( )$ is the gamma function.
%(its definition can be found, for ex., in \cite{ } ).

We also define
 a probability distribution $\{\omega =
\omega_1, \omega_2, ... \}$ on integers $\{ 1, 2, ... \}$ by
\begin{equation}\label{om} \omega_1 = 1 - 1/ \log 3,\: ... \,,\:
\omega_i\,= 1/ \log (i+1) - 1/ \log (i+ 2),\: ... \; ,
\end{equation}
$i= 1, 2, ... \, .$
(In what follows we will use this distribution, but
the theorem described below is  true for   any
distribution with nonzero probabilities.)

The measure $R$ is defined as follows %%%d
\begin{equation}\label{R}
R( x_1...x_t ) =  \sum_{i=0}^\infty \omega_{i+1} K_i ( x_1...x_t ) \, .
\end{equation}
It is important to note that the measure $R$  is a
universal measure  for the class of all stationary and
ergodic processes with a finite alphabet; Ryabko (1984). Hence, $R$ can be used as  a consistent estimator of probabilities.

Let us describe the scheme of the forecasting method based on the measure $R$ for the sequences
generated by the sources of different types.
\subsection{Finite-alphabet case}

As we mentioned above the  measure $R$  can be applied for prediction. More precisely we may use $R$
for defining the following conditional probability as:
$$
R(a | x_1...x_t ) =  R( x_1...x_t a) /
R( x_1...x_t )  \, ,
$$
$a \in A$.
In the finite-alphabet case the scheme of the prediction algorithm is quite simple. Let $x_1 ... x_t$ be a
given sequence. For each $a \in A$
%( $A$ is the set of all different $x_i$ from the sequence $x_1 ... x_t$ )
we construct the sequence $x_1 ...x_t a $ and compute the value $R(a | x_1...x_t )$ . Having the set of such conditional
probabilities we use them as   estimations of the unknown probabilities $P(a | x_1 ...x_t )$, $a \in A$.
\subsection{Real-valued case}
Let $(\Omega, F , P)$ be a probability space and let $X_1 , X_2
,...$ be a stochastic process with each $X_t$ taking values in a
standard Borel space. %%%d хрен ли он Борель когда он просто R
 Suppose that the joint distribution $P_n$
for $( X_1 , X_2 ,..., X_n )$ has a probability density function
$p( x_1 x_2 ...x_n )$ with respect to the Lebesgue measure $L$.
(A more general case is considered by Ryabko (2009). In particular, it is
shown that any universal measure can be used instead of $R$.)  Let
$\Pi_n$,  $n \ge 1$ , be an increasing sequence of finite
partitions $\Omega$ that asymptotically generates the Borel
sigma-field $F$, and let
 $x^{[k ]}$ denote the element of $\Pi_k$ that contains the point $x$. For integers $s$ and $n$ we define the following
approximation of the density:
$$
p^s ( x_1 ,..., x_n ) = P( x_1^{[s ]} ,..., x_n^{[s ]} ) / L (  x_1^{[s ]} ,..., x_n^{[s ]} ) \, .
$$
 Now we define the corresponding density $r$ as follows:
\begin{equation}\label{r}
r( x_1...x_t ) =   \sum_{s=1}^\infty \omega_{s} R( x_1^{[s ]} ,..., x_t^{[s ]} ) / L (  x_1^{[s ]} ,..., x_t^{[s ]} ))
\end{equation}
It is shown by Ryabko (2009) that the density $r ( x_1 ...x_t )$ estimates the unknown density $p( x_1 ,..., x_t )$ , and the
conditional density
\begin{equation}\label{dr}
 r(a | x_1...x_t ) = r (  x_1...x_t a) / r (  x_1...x_t)
\end{equation}
is a reasonable estimation of $p(a | x_1...x_t )$.

In practical application of this method we may face a question about choosing the length of the input
data to be considered. One may think that  the best result will be achieved when using the
maximally available quantity of known time series elements. However, it is not always true, because
actually statistical characteristics of the process may vary in time 
(or the process may have a large memory), %%%d  large memory это наоборот противоположный случай
 and then ``old'' data contain no information on ``new'' statistical characteristics. So, if the answer is not obvious, to estimate
 the probability one should use $\bar{r}$ measure defined below.
Let $x_1...x_t$ be a given sequence. Consider next a set of
distinct natural numbers $N$. For each $n$ from $N$ consider the
sample $x_{t-n+1} ...x_t$. In other words, in fact, each sample
consists of $n$ last elements of the original time series. Each sample taken
individually  will be called a  ``window'' with appropriate $n$ %%%d
as its size. Suppose $N = $ $\{ n_1, ... ,n_k \} $, $k \ge 1$,  then
$$
\bar{r}(x_1...x_t) = \sum_{i=1} \bar{\omega}_i r(x_{t-n_i+1} ...x_t)
$$
where $\bar{\omega}_1, \bar{\omega}_2, $ $ ..., \bar{\omega}_k$ is
a probability distribution on $N=\{ n_1, ... ,n_k \}$. (In our
calculations we use the following distribution: $ \bar{\omega}_i =
\omega_i, $ $i = 1, ... , k-1$ from (\ref{om}), and
$\bar{\omega}_k = 1/ \log (k+1)$.) As well as $r$ we may use
$\bar{r}$ for estimation of conditional probability $p(a |
x_1...x_t ) $. As seen from the definition, $\bar{r}$ is a
``mixture'' of probabilities for the appropriate time series of
different lengths. Thus estimations based on $\bar{r}$ seem to
automatically provide close to optimal results in the context of usage
simplicity and precision of forecast. In what follows, such  way of
calculations will be called adaptive mode.

It is worth noting, that such a ``mixture'' was already used several times
in this  work (see  (\ref{R}) and  (\ref{r}) ). The point is that
the performance of such ``mixtures'' is close to the performance
of the best summand (see details and theoretical justification in
(Ryabko, 2010).

\subsection{Implementation of the method}
Consider next some aspects of the implementation of the investigated method.
Suppose there is a certain source which generates values from some
real-valued interval $ [A; B] $ and we have time series
$x_1...x_t$ generated by this source. %%%d
 The next   value $x_{t+1}$ needs to be
predicted. For the purpose of simplicity we will consider 
computations based on $r$. The case of $\bar{r}$ is analogous. %%%d

{\it Step 1.} Calculate $ r ( x_1...x_t )$ using (\ref{r}). We will describe this step in more details. Divide the
interval $[A; B]$ into two equal partitions, called bins, %%%d везде замени bucket на bin
 and transform $x_1...x_t$ into a sequence of
 symbols each of which is equal to the index of the bin, that contains the appropriate point $x_i$. (I.e. if
0 and 1 are the bins indexes, we will then obtain from $x_1...x_t$  the sequence consisting only of these
symbols). Then calculate the first item of the sum from the formula (\ref{r}). As the value of  $L$ we take a product of
lengths of all buckets containing $x_i$ , and the  measure $R$ is computed for the sequence obtained as a
result of transforming $x_1...x_t$ for the current quantization. After that again we divide each of the existing
bins (there are two) into two equal bins (there will be four) and for this new quantization do
analogous operations to compute  the second term from the right-hand side of the equation (\ref{r}). Go on
in this way until getting the quantization  for which each of the distinct time series values (including
those added at the next step) belongs to  distinct bins. %%%d
 Summing all  terms,  obtain $r ( x_1...x_t )$ from
(\ref{r}). It should me mentioned that at this step any other algorithm for achieving the increasing
sequence of finite partitions of $[A; B]$ may be applied.

{\it Step 2.} Consider the set $P$ consisting of points $A$,$ A + h$, $A + 2 h,$ $..., B$ , where $h$ is a certain small %%%d
 constant. (In this paper  $h=0.01$ was used.) For each element $a$ of this set we construct the sequence
$x_1 ...x_t a$  and compute the value $r ( x_1...x_t a)$ similarly. %, where а is an element from the set P.
Here we use the quantizing %%%d vezde zameni  quantizing na quantization
 that is the same as for previous step.
Then  by formula (\ref{dr}) for each element from the set $P$
calculate the estimations of the appropriate conditional
probabilities and find the corresponding prediction. In
this paper the forecast value was considered to be equal to the
element from $P$ with the biggest estimate of conditional
probability. But any other adequate approach may be used.

\section{ Experimental results}
In this section the results of experimental estimation of the method, described in the previous
section, are given. All the  processes investigated  had values from the corresponding interval. %%%d kakoi interval??
 To have a possibility to compare the results obtained by the investigated method, with the
corresponding actual values, all time series values used in the experiments were taken from the 
past. %%%d eto ochevidno
The computations in this section can be divided into the two independent
parts. In the first part %%%d kakaya eshe branch
 we considered   only the forecasting method, while in %%%d
the second part some preprocessing was used.  %%%d %%%d
The detailed descriptions of the computations, %%%d???? 
 along with the results of these
experiments are given below, in the corresponding subsections.

{\it Simple time series forecasting.}
As the target  here we chose the time series consisting
of the following indices: monthly and smoothed monthly means of
sunspot numbers, absolute daily and monthly solar flux values. All
datasets used in the experiments of this subsection, can be found
at the National Geophysical Data Center (NGDC) Internet site in
the ``Space Weather $\&$ Solar Events'' section. This
subsection includes the study of two items.

{\it One-step ahead forecasting.} The content of  each
experiment   %%%d eto mne neponyantno, kakoi item ??? 
can be described as follows. Given $t$
successive values of certain time series, we tried to forecast
its $(t+1)$th element. For the purpose of the analysis of connection
between the known values and the forecast precision for
each process the time series with different lengths were
considered. Moreover, in order to explore the possibility of
computational optimization,  additional calculation in adaptive
mode was carried out. %%%d hren kakayato
(See previous section for this method
details). We should mention here, that the set of ``window'' sizes
for each process in this case was the same as that consisting of
all length used for this time series during the calculations %%%d eto ya vooblshe ne ponyal
without the ``window'', and each item of the sum was equiprobable
in the ``mixture''. For each length of the time series related to a
certain process, and also for the corresponding calculations in
adaptive mode, there were 25 experiments on  independent
datasets. After  doing all appropriate computations,
the precision estimation was made by considering the differences
between the  forecast and actual values. The results of
the experimental computations of this stage are given in Table
1 below.  The first column contains the name of the
investigated time series, the second gives the range of its
values. The rest of the columns contain the mean absolute error (MAE),
obtained either in the calculations in the adaptive mode, or when
using the corresponding length of the time series. The ``n/d''
(``no data'') text means that no calculations were carried out,
because  there was no data. %%%d
For example, the information in the third
row of the table indicates that for the time series on
 absolute daily solar flux the MAE corresponding to the
calculations based on the 4000 known values, is 1.45. All time
series values belong to the range [50; 300]. Also, taking into
account the first row of the table, we may say that for the monthly sunspot numbers 
data  the MAE for
calculations in adaptive mode, using the ``window'' with sizes
$500, 700, 1000, 1200, 2000 $ and $3000$, is $23.97$. %%%d

\begin{table}[ht]

 \center
 \label{tab1}
 \caption{Experimental computations results of the subsection  3.1}
 \vspace{2mm}
 \newcolumntype{Y}{>{\centering\arraybackslash}X}
 \newcolumntype{Z}{>{}m{34mm}}
 \newcolumntype{Q}{>{\centering\arraybackslash}m{18mm}}
 \newcolumntype{A}{@{}Y@{}}

 {
  \renewcommand{\tabularxcolumn}[1] {m{#1}}
  \begin{tabularx}{165mm}{|@{\hspace{1mm}}Z@{\hspace{1mm}}|@{}Q@{}|A|A|A|A|A|A|A|A|}
  \hline
  \multicolumn{1}{|c|}{\multirow{2}{*}{Time series}} & \multirow{2}{*}{Range}   & \multirow{2}{*} {Adapt.} &
  \multicolumn{7}{c|}{Length}  \\
  \cline{4-10}
      & & & 500 & 700 & 1000 & 1200 & 2000 & 3000 & 4000       \\
  \hline
  { Monthly sunspot number means} &  [0; 256] & 23,97 & 6,54 & 2,56 & 9,58 & 15,85 &21,7 & 19,63 & n/d \\
  \hline
  Smoothed monthly sunspot number means& [0;210] &2,34 & 1,5 & 1,1 & 1,99 & 0,77 & 3,36 & 2,56 & n/d\\
  \hline
  Absolute daily solar flux  & [50;300] & 1,78 & 1,17 & 1,17 & 2,71 & 5,52 &8,35 & 1,72& 1,45 \\
  \hline
 Absolute monthly solar flux&[580;2540]& 45,29 & 211,29 & 45,88& 2,71 & n/d &n/d & n/d& n/d\\
 \hline

 \end{tabularx}
 }
\end{table}

{\it  Short-dated forecasts comparison.} $\,$ Here we compared
the short-dated forecast for smoothed sunspot numbers, which was
published at the NGDC site, with that obtained from the method
based on $R$ measure. According to the information, provided at
the NGDC site today there exists a special forecasting software
(in what follows - NGDC software), which forms a ``preliminary
look'' at the current solar cycle using the improved
McNish-Lincoln method. In accordance with the information
published on the server, NGDC software constructs the forecast for
smoothed monthly sunspot number means and also uses the known
values of the time series elements for the certain process. As the
NGDC software forecast the file {\it sunspot.predict} of
06.08.2008 published on the site was taken, where for each month
of the current solar cycle the corresponding forecast value and
its confidence interval are given. We should note here that in our
comparison we considered just the forecast value, without taking
into the account the confidence interval. It is known that at the
forecast producing moment for the process there were available and
consequently usable data up to February, 2008. Hence, at the
computations carrying out moment in June of 2010 there were 21
already known values of the investigated time series (for each
month from March, 2008 till November, 2009). Computations by the
explored method without ``window'' usage were carried out
according to the following scheme. At first, we point out the time
series having the defined length within the certain experiment and
with its last element related to February, 2008. Basing on the
sequence, we get the one-step ahead forecast, i.e. prediction on
March, 2009. Then, at the next iteration we shift the borders of
the time series with known values to the right on the one
position, replacing the new last sequence element (related to the
actual value of March, 2009) with the forecast value obtained at
the previous iteration. Basing on this changed time series, we
again get the one step ahead forecast, on April, 2008. And so on.
Finally, for each observed length of time series 21 experimental
computations were carried out. The last forecast value was
achieved for November, 2009. The calculations in adaptive mode
were accomplished in a similar way but with observing the set of
time series of different lengths at the each iteration instead of
considering one time series. Likewise in the previous item, the
set of ``window'' sizes for each process in this case was the same
as that, consisting of all lengths used for this time series
during the calculations without the ``window'', and each item of
the sum was equiprobable in ``mixture''. After the accomplishing
of all appropriate computations, the precision estimation was made
by considering the differences between corresponding forecast and
actual values. The results of experimental computations for the
investigated method are placed in Table 2. The first row contains
MAE for $R$ measure based method and the appropriate input data.
The second row includes MAE for computations, applying the NGDC
software. So, it is seen, from the table that for the forecast
using the described method and the time series of size 1000 the
MAE is 1.87 for 21 experiments. For the NGDC software this value
was calculated on the basis of one forecast file.

For illustration purposes the results obtained at this stage of
experimental estimation are shown at the Fig. 1 below. At the
graphic $N$ axis is the smoothed monthly sunspot numbers and $t$
axis presents the time units with each related to the certain
month of the solar cycle. The part of the graphic with $t \le 0$
is the period with data for forecast producing, and the part with
$ 1 \le t \le 21$ is the period of time for which the forecasts
were made. The solid bold line displays the actual process values
(note that they are from the end of time series only), and the
dashed lines with short and long dashes plot the forecast values
achieved by NGDC software and by the method based on $R$ measure
in adaptive mode correspondingly. In the case of using NGDC
software the comparable MAE was observed only for the first four
forecast values and was 1.95. After the ten experiments it was
growing up to 5.51 and, finally, after all 21 computations of this
number of experiments the value equaled 16.39. So, we may say,
that there was obvious trend of the MAE increasing with the
growing of the forecast index number.

% \begin{figure}   \label{ris2} \centering \includegraphics[scale=1]{pic.bmp}      \end{figure}

\begin{figure}
  \center
  \includegraphics[width=15cm, keepaspectratio]{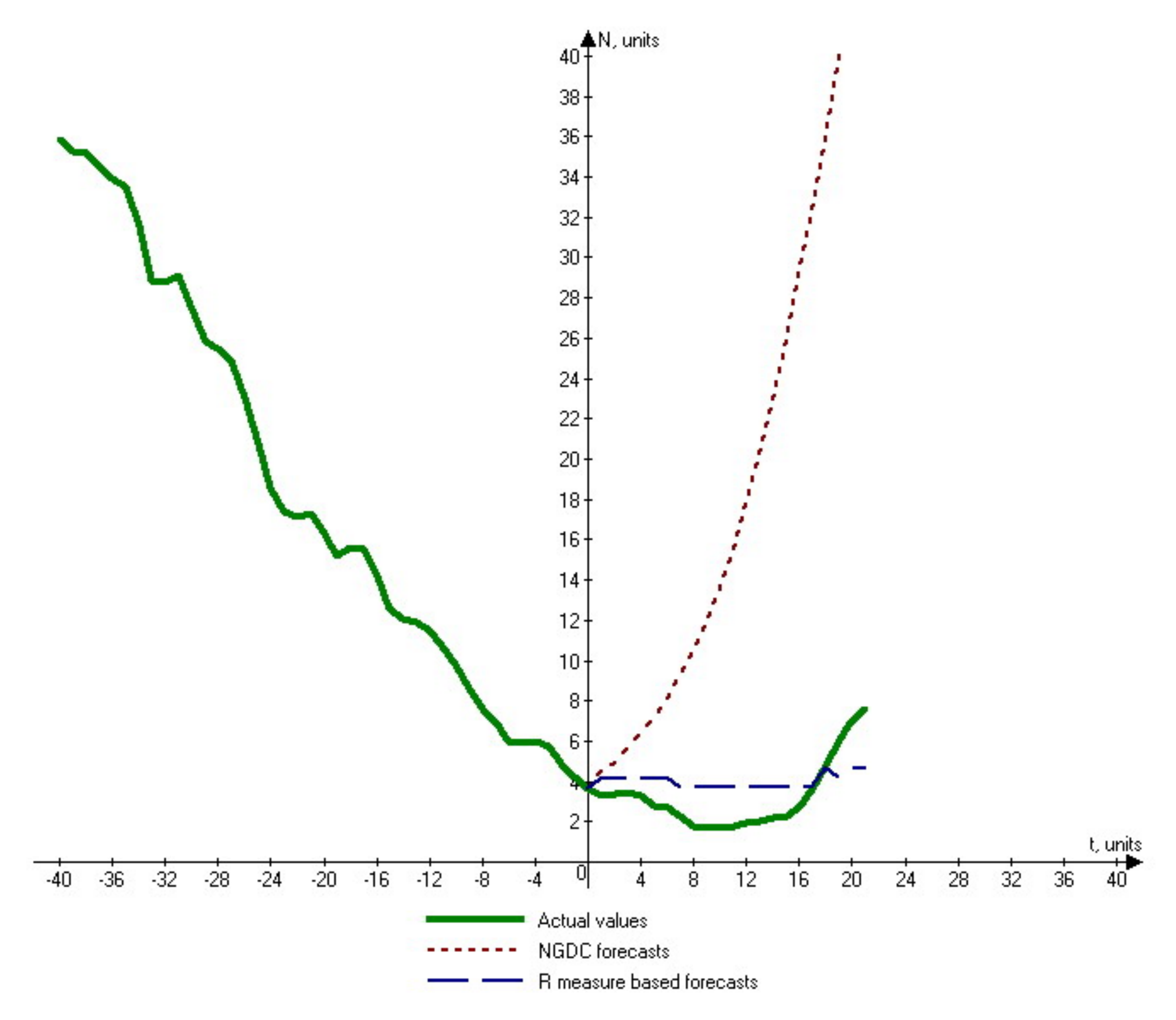}
  \caption{Results of short-dated forecasts comparison}
\end{figure}

Therefore, the graphic shows that in this case the forecasting
method based on R measure was more effective in the producing of
the short-dated prediction. There is no information on the NGDC
server to carry out other experiments for precision comparison. It
goes without saying, that the comparison of forecasting methods
with only on file to take into account may be not enough to come
to the ultimate conclusions, however, any way the rate of the
absolute and relative (we may see that it is about $1\%$) errors,
obtained by the investigated method, justifies its right of
existence. 

{\it Forecasting with preparatory time series transformation.} It is known that statistical methods of forecasting
are often used with the preparatory transformation of input data.
The combination of this approach and the investigated method may
improve the precision of prediction. In this subsection
experimental estimation of the application of the forecasting
method based on the measure $R$ together with the preparatory
differentiation of the time series was accomplished. Having such
forecast value we can easily add it to the last known elements of
time series and obtain the desired prediction Again the subsection
includes the study of two items.

{\it  One-step ahead forecasts comparison. } Here as the target
of research time series, consisting of the 15-minute sea level
indices, were considered. All datasets used in the experiments of
this item, can be found at the British Oceanographic Data Centre
(BODC) Internet site in the ``UK Tide Gauge Network'' section. The web interface of the site allows any registered user
to download data file with the history of 15-minute sea level
indices for the certain gauge from some location. In this file for
each timestamp for the certain period of time the appropriate
actual and residual values are given. The latter are calculated
from the observed sea level values minus the predicted sea level
values. According to the provided by BODC support information
predicted tide values are produced at the National Oceanography
Centre's (NOC), using their harmonic tidal analysis. This is based
on the TIRA tidal analysis programs following the Doodson method.
We considered the data related to the Bubbler tide gauge from
Dover and captured in 2005 year, because all values for that
period are available and there are no missed or interpolated
points there. As in the first item of previous subsection the
one-step ahead forecasting was decided here. Furthermore, the
scheme of the calculations, using the time series of different
length and adaptive mode, was the same, but in this item each
estimation series contained 30 experiments. After the
accomplishing of all appropriate computations, the precision
estimation was made by considering the differences between
corresponding forecast and actual values. As we mentioned above
the history contains residuals for every element of time series.
So, in the purpose of the objectivity for each MAE of NOC software
we took into account residuals corresponding only to the values
predicted by investigated method. The results of the study of this
item are situated in Table 3.

\begin{table}[ht]
  \center
  \label{tab2}
  \caption{ Forecasting results 15-minutes sea level indices}
  \vspace{2mm}
  \begin{tabular}{|l|c|c|c|c|c|}
   \hline
   \multicolumn{1}{|c|}{Length} & 500 & 1000 & 2000 & 5000& Adaptive mode \\ \hline
   $R$ & 0.034 & 0.038 & 0.0396 & 0.037 & 0.037 \\ \hline
   NOC & 0.207 & 0.2133 & 0.09883 & 0.0779 &  $-$ \\ \hline
  \end{tabular}
\end{table}

The first row contains MAE for $R$ measure based method and the
appropriate input data. The second row includes MAE for
computations produced at the NOC. So it is seen from the table
that for the forecast using the described method and the time
series of size 5000 the MAE is 0.037 for 30 experiments. And if we
consider the given by NOC residuals for each of that 30 predicted
by the investigated method values we will obtain 0.0779. There is
no information about calculation in appropriate adaptive mode for
NOC.

{\it  Comparison with the simplest method } The simplest test to
explore a new forecasting method is a comparison with so-called
inertial prediction, where the last actual value is supposed to be
the next forecast value. Here we will accomplish this kind of
comparison, regarding the FOREX cross-rates daily currencies of US
dollar and Great Britain pound to Euro. We used data on the both
cross-rates from January 2001, 03 to January 2011, 17. All
datasets used in this investigation item were taken from  FXHISTORICALDATA.COM.
Again we considered the one-step ahead forecasting. Each
estimation series contained 10 experiments. After the
accomplishing of all appropriate computations, the precision
estimation was made by considering the differences between
corresponding forecast and actual values. The comparison results
are summarized in Table 4 and Table 5.

\begin{table}[ht]\
     \center
     \label{tab3}
     \caption{ Comparison with the inertial method for USD/EUR cross-rate}
     \vspace{2mm}
     \begin{tabular}{|l|c|c|}
     \hline
  \multicolumn{1}{|c|}{Foracasting method} & 500 & 1000 \\ \hline
$R$ & 0.00298 & 0.00582 \\ \hline
Inertial & 0.0030 & 0.0059 \\ \hline
\end{tabular}
\end{table}

\begin{table}[ht]
     \center
     \label{tab4}
     \caption{ Comparison with the inertial method for GBP/EUR  cross-rate }
     \vspace{2mm}
 \begin{tabular}{|l|c|c|}
\hline
 \multicolumn{1}{|c|}{Foracasting method}& 500 & 1000 \\ \hline
$R$ & 0.0019& 0.00089\\ \hline Inertial & 0.0025 & 0.0010 \\
\hline
\end{tabular}
\end{table}

Tables have similar structure. The first and the second rows contain the MAE for forecasting
method based on $R$ measure and inertial prediction respectively when using the appropriate length of
input data. As a whole we may say that the forecasting method based on the universal measure $R$
showed the better than the inertial prediction results.

\section{ Conclusion}
In this article the implementation and experimental estimation of
forecasting method, based on the universal measure $R$ were
considered. Analysis of the investigation outcomes has shown the
quite high precision of the obtained results. Moreover, in
comparison of the short-dated forecasts, obtained by the $R$ measure
based method and by NGDC software the significant advantage of the
former was detected. Good results were also achieved in the
combine with preparatory transformation of time series. We found
parameters for which the consistent superiority of the considered
method in comparison with UK NOC prediction was detected in the
one-step ahead forecasting. As a result we may conclude that
universal codes are believed to be the effective tool for the
forecasting methods construction in practical application.


\begin{thebibliography}{1}

\bibitem{GH}
 De Gooijer, J.G. \& Hyndman, R.J. (2006)
 25 years of time series forecasting.
 {\it International Journal of Forecasting}, 
 22(3),  443-473 



\bibitem{Fi}
Fitingof, B.M. (1966). Optimal encoding for unknown and changing statistics
of messages. {\it Problems of Information Transmission}, 2(2), 3-11.

\bibitem{FX}
 FXHISTORICALDATA.COM: URL: http://www.fxhistoricaldata.com/



\bibitem{Ga}
Gallager,  R.G.  (1968). {\it Information Theory and Reliable Communication}. John
Wiley \& Sons, New York.


\bibitem{K}
Kolmogorov, A.N.  (1965). Three approaches to the quantitative
definition of information. {\it Problems Inform. Transmission}, 1,
3-11.

\bibitem{Kr}
Krichevsky, R.  (1993). {\it Universal Compression and Retrival}, Kluver
Academic Publishers,


\bibitem{MVH}
Makridakis, S.G.,  Wheelwright  S.C., and Hyndman, R.J. (1998).
  {\it Forecasting: Methods and Applications}, Wiley,  (3rd edition).



\bibitem{Ri}
Rissanen, J. (1984).  Universal coding, information, prediction, and
estimation. {\it IEEE Trans. Inform. Theory}, 30(4), 629-636.


\bibitem{Ry0}
 Ryabko, B. (1984). Twice-universal coding. {\it Problems of
Information Transmission}, 20(3), 173-177.



\bibitem{Ry1}
Ryabko, B. (1988). Prediction of random sequences and universal
coding. {\it Problems of Inform. Transmission}, 24(2), 87-96.


\bibitem{Ry09}
Ryabko, B. (2009). Compression-Based Methods for Nonparametric Prediction and Estimation of Some Characteristics of Time Series. 
{\it IEEE Transactions on Information Theory}, 55(9), 4309-4315. 

\bibitem{Ry10}
 Ryabko, B. (2010). Applications of Universal Source Coding to Statistical Analysis of Time Series.
  In: Isaac Woungang et al. (Eds.), {\it Selected Topics in Information and Coding Theory},
   World Scientific Publishing, pp. 289 - 338. (see also http://arxiv.org/pdf/cs/0701036v2 ) 

\bibitem{RA06}
Ryabko, B. \& Astola, J.  (2006) Universal Codes as a Basis for Time Series Testing
 {\it  Statistical Methodology}, 3, 375-397.


\bibitem{RM}
 Ryabko,  B. \&  Monarev, V. (2005). Using Information Theory Approach to  Randomness
 Testing.  {\it
Journal of Statistical Planning and Inference}, 133(1),   95-110



\bibitem{SW}
 Space Weather \& Solar Events. National Geophysical Data Center: URL:
http://www.ngdc.noaa.gov/stp/spaceweather.html


\bibitem{UN}
 UK National Tide Gauge Network. British Oceanographic Data Centre: $ URL:
https://www.bodc.ac.uk/data/online_delivery/ntslf/ $








\end{thebibliography}
\end{document}